# Structure-function relationships of fullerene esters in polymer solar cells: unexpected structural effects on lifetime and efficiency

Michael Tro[1], Alexis Sarabia[1], Kyle J. Bandaccari[1], David Oparko[2], Emma Lewis[2,3], Maxwell J. Giammona[1], Justin Isaac[2], Parisa Tajalli-Tehrani Valverde[1], Grace E. Chesmore[2,3], Thorsteinn Adalsteinsson[1,3], Richard P. Barber, Jr.[2,3] and Brian J. McNelis[1*]

[1]*Department of Chemistry and Biochemistry, Santa Clara University, Santa Clara, CA 95053, USA*

[2]*Department of Physics, Santa Clara University, Santa Clara, CA 95053, USA*

[3]*Center for Nanostructures, Santa Clara University, Santa Clara, CA 95053, USA*

*Corresponding author.Tel.:+1 408 554 4797; Fax:+1 408 554 7811.

E-mail address: bmcnelis@scu.edu (B.J. McNelis).

## Abstract

We report both transport measurements and spectroscopic data of polymer/fullerene blend photovoltaics using a small library of fullerene esters to correlate device properties with a range of functionality and structural diversity of the ester substituent.  We observe that minor structural changes can lead to significant and surprising differences in device efficiency and lifetime. For example we have found that isomeric R-groups in the fullerene ester-based devices we have studied have dramatically different efficiencies. The characteristic lifetimes derived from both transport and spectroscopic measurements are generally comparable, however some more rapid effects in specific fullerene esters are not observed spectroscopically.  It is apparent from our results that each fullerene derivative requires re-optimization to reveal the best device performance.  Furthermore we conclude that a library approach is essential for evaluating the effects of structural differences in the constituent molecules and serves as important device optimization method that is not being currently employed in photovoltaic investigations.

**Introduction**

Third generation solar cells (solution processable) continue to grow as an area of research given their promise as an inexpensive alternative to silicon technologies [1,2]. The cost benefits include easy fabrication and elimination of an expensive raw material commodity [1–13]. Recent among this newer class of solar cells are the perovskites [14,15]. Although they exhibit power conversion efficiencies in the 20% range [16], they also have disadvantages: toxic metal content and instability in the presence of humidity [15]. Polymer-fullerene blend photovoltaics also have two major obstacles to implementation: relatively poor power conversion efficiencies ($\eta$'s) and short device lifetime. Given the current state of third generation solar cell development, it is apparent that continued effort on these multiple fronts is needed to advance these systems to a viable product stage.

To date the bulk of research on the polymer blend systems has focused on device fabrication and the synthesis and modification of polymers to improve efficiencies, which have been optimized at 9% or higher [4,5,12,13]. In contrast, our target has been the characterization of solar cells as we modify the fullerenes, with close attention to both efficiency and stability. A widely studied organic photovoltaic system utilizes a [6,6]-phenyl-$C_{61}$-butyric acid methyl ester/poly(3-hexylthiophene-2,5-diyl) blend, or PCBM:P3HT, as the electron acceptor and photon induced electron donor respectively (Fig. 1). PCBM, a functionalized $C_{60}$, is used to improve solubility since the unmodified version tends to form clusters and aggregates [17–21].

Other esters of the [6,6]-phenyl-C$_{61}$-butyric acid are commonly abbreviated as PCB-alkyl derivatives. In previous experiments in our laboratories alternative electron acceptors were synthesized including [6,6]-Phenyl C$_{61}$ butyric acid octadecyl ester (PCBOD, **1a**) [22] and [6,6]-phenyl C$_{61}$ butyric acid octyl ester (PCBO, **1b**) [23] (Fig. 1). Based on detailed results for PCBOD and preliminary ones for PCBO, we have produced a more diverse set of PCB-alkyl esters and measured the figures of merit for photovoltaic devices using these additives. The purpose of this study is to demonstrate that even small structural changes in constituents can have a dramatic affect on device performance. Most researchers in this field limit their approaches to "rational" design of device components with little recognition in the literature that device self-organization is poorly understood and can be easily perturbed through structural changes of the constituents. We believe that to properly explore device parameter space, the library approach can yield surprising differences in device performance. We specifically chose sets of compounds to prepare that on first approximation should yield almost no differences in the device transport measurements. Although our devices are not high-performing, the purpose of our study is to prove the value of the library approach and demonstrate that performance increases and decreases can occur based on "random" structural changes of components. Even if this approach yields only 5-10% improvements in device performance in state-of-the-art devices, that increase would be significant and could translate to commercial viability.

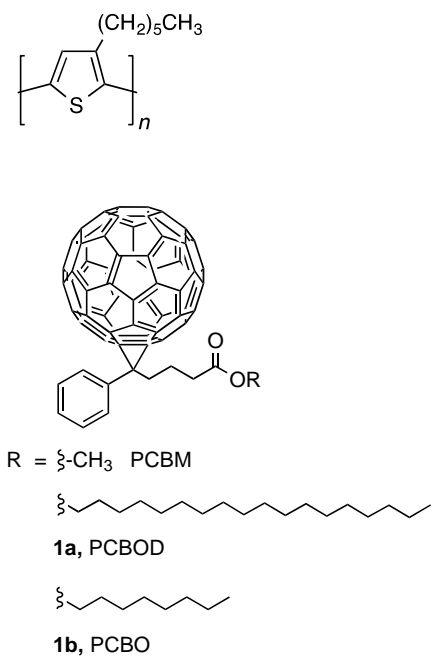

R = §-CH₃  PCBM

**1a,** PCBOD

**1b,** PCBO

Fig. 1. Electron-donor and two initial electron-acceptor molecules used in this study.

We characterize blend stoichiometry of our samples with $C_{60}$ mole fraction $x$ where

$$x = \frac{m_{C60}/MW_{C60}}{m_{C60}/MW_{C60} + m_{P3HT}/MW_{P3HT}}$$

with *MW* denoting the molecular weight of each species, *m* denoting its mass and $C_{60}$ referring to the functionalized fullerene molecule under investigation. For the electron donor (in this expression P3HT), we use the monomer molar mass. In other words, *x=0.5* represents a blend with one $C_{60}$ molecule per P3HT monomer [22–24]. Notation in the literature often uses weight concentrations. The commonly studied equal weight blend corresponds to a PCBM mole fraction *x=0.16*. We have chosen to use mole fraction as our primary notation given that we compare fullerene-polymer blends using *different* fullerenes with *different* molecular weights. This approach allows us to report our results with a simple ratio of fullerene-to-monomer units.

In our solar cells studies of PCBOD **1a** and PCBO **1b**, we observe a fourfold increase in power conversion efficiency for PCBO over PCBOD, albeit at 140ºC anneal temperatures [23]. For both PCBM and PCBOD, we observed the best efficiency and lifetime with 195ºC anneals: a temperature just below the melting point for P3HT [25]. Before our initial PCBO results, a reasonable conclusion would have been that the optimum anneal temperature is a characteristic dictated by the polymer. The significantly lower anneal temperature for the PCBO suggests that this simple picture is inadequate and that a more detailed investigation is needed. These findings motivated us to produce a small library of fullerenes.

Combinatorial approaches have been reported recently to optimize parameters of polymer photovoltaics [26,27]. In one study, novel MEH-PPV analogs were prepared and the fullerene structure analyzed for optimal device performance [28]. They found that with different conducting polymers there are specific fullerene structures that maximize efficiency with each polymer they studied. These *paired* polymer-fullerene combinations demonstrate that changing components of the blend requires re-optimization of fabrication and constituent structure selection.

We prepared simple alky1 chain esters and some functionalized side chains to examine the device lifetime and efficiency vs. R-group chain length and functionality. We use a simple transesterification method to prepare the fullerene esters **1c-i** as shown in Scheme 1 [29]. Following we present efficiency and lifetime results for these new fullerenes in addition to a more complete data set for PCBO (not previously reported).

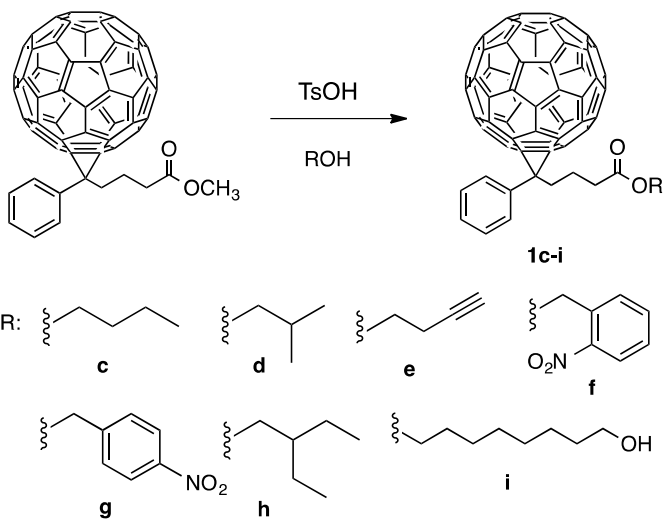

Scheme 1: Synthesis of Fullerene Esters **1a-i**

**Experimental**

Sample fabrication is described in previous work [22] and briefly summarized here. Indium-tin-oxide (ITO) coated glass substrates are patterned before wet-processing steps (spin-coating and annealing) are performed in an MBraun inert atmosphere glove box. The two spin-cast and annealed layers are poly[3,4-ethylenedioxythiophene]:poly[styrenesulfonate] (PEDOT:PSS) followed by the active layer blend. The final electron-injecting contacts are evaporated in a separate bell jar evaporator using 2 nm thick lithium-fluoride (LiF) followed by 100 nm of aluminum (Al). Samples are carried from the glove box to the evaporator in a vacuum tight vessel containing glove box nitrogen, and transferred and evacuated in less than 5 minutes.

Current-voltage (*I-V*) characteristics of the devices are measured in ambient atmosphere in darkness and under illumination by a PV Measurements, Inc. Small-Area Class-B AM1.5 with a Keithley 2400 SourceMeter. A Varian Cary 50 spectrometer is used for UV-vis measurements. Both spectroscopic and transport experiments are carried out for periods of hours or days. Fig. 2 shows a schematic of the device architecture for electrical transport measurements. Spectroscopic studies were performed on samples directly from the glove box with no ITO or Al contacts.

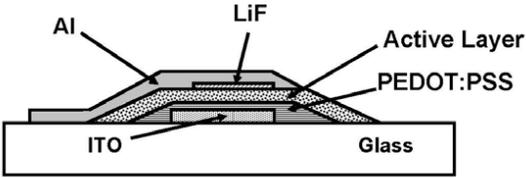

Fig. 2: Schematic of solar cell layers.

**Results and Discussion**

Fig 3 presents typical transport curves used for deriving figures of merit. *I-V* sweeps are recorded at fixed time intervals (usually 15 min). Many curves have been removed from the figure for clarity. This particular sample was a PCBO blend, however all the various samples yielded similar results, albeit at differing current scales. Efficiency, open circuit voltage and short circuit current density ($\eta$, $V_{OC}$ and $J_{SC}$) are calculated from each sweep. *η(t)* appears linear in a semilog plot (inset), so it can be written as

$$\eta(t) = \eta(0)\exp(-t/\tau),$$

where a characteristic (1/e) time $\tau$ can be derived from the slope (*-1/τ*) of the fitted line [22]. We define this $\tau$ as the characteristic "lifetime" for our sample.

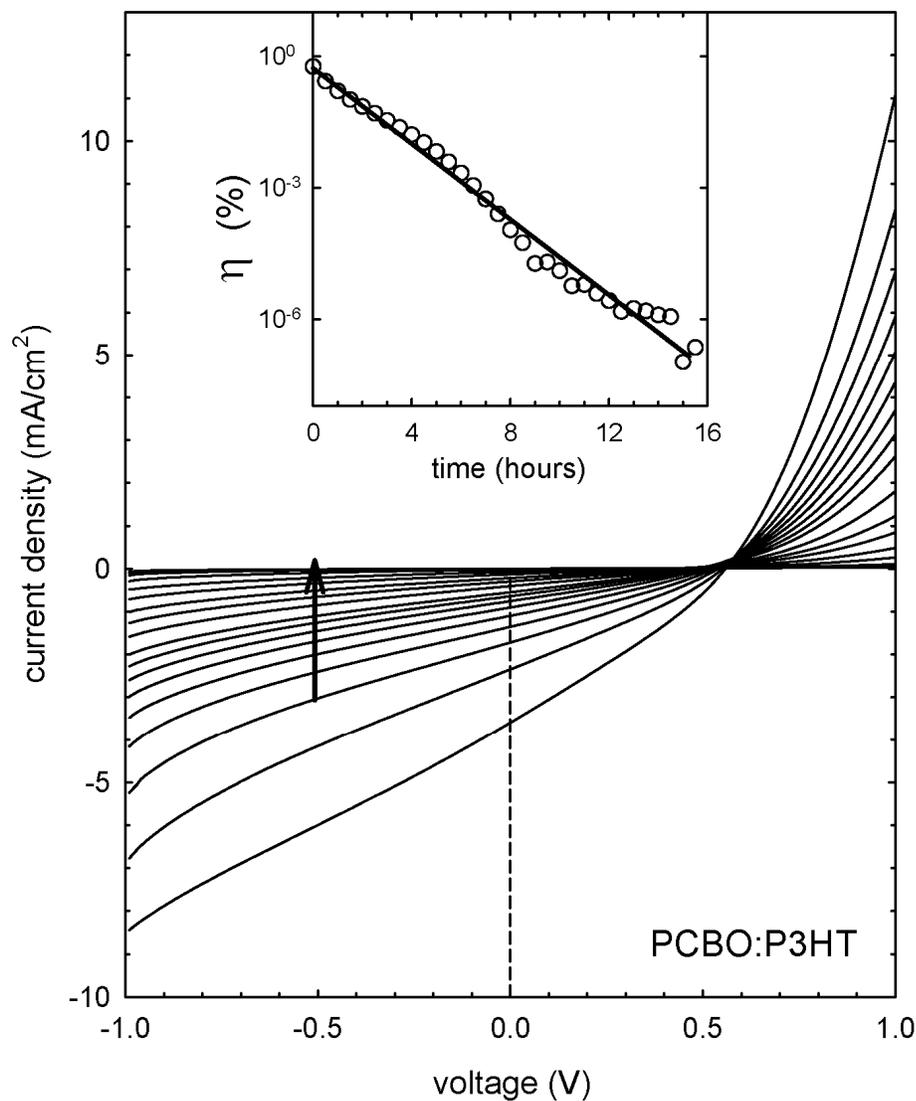

Fig. 3. Typical current-voltage (*I-V*) curves showing the degradation of a PCBO:P3HT device in ambient conditions. The arrow indicates the progression of time. *Inset:* a semilog plot of the power conversion efficiency of this device as a function of time in ambient conditions. The solid line fit shows the slope used to extract the characteristic time $\tau$.

Fig. 4 summarizes our results for PCBO:P3HT devices. As previously observed in the preliminary data, samples annealed at 140 °C were superior to those annealed at 195 °C. Our previous experiments had shown that for both PCBM and PCBOD (octadecyl chain) samples show the best efficiency and lifetime after being annealed at 195 ºC. It is appealing to conclude that since these rather different PCB-esters are optimized at the same anneal temperature; the polymer is the constituent that controls this condition. The melting temperature for P3HT is given as 218 ºC [25]. The fact that our PCBO samples do not optimize at this temperature indicate that there is more complexity to this aspect than originally understood. That this result persisted after a more thorough investigation motivated us to produce and study a broader array of PCB-esters. We have now further explored the structure-function relationships of various fullerene esters (shown in scheme 1) and the effects of R-group changes on device efficiency and lifetime.

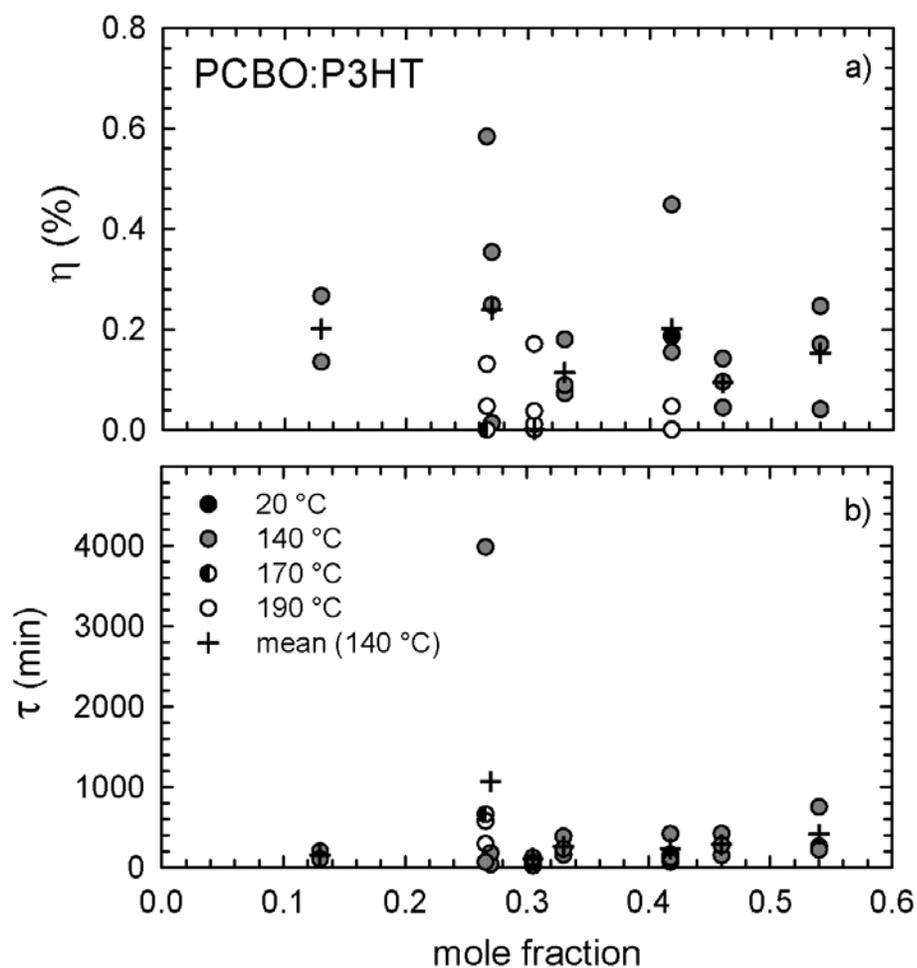

Fig. 4. Efficiency and lifetime and data from a series of PCBO:P3HT devices. We plot a) the initial power conversion efficiency $\eta$ and b) the decay time constant $\tau$ as a function of PCBO mole fraction (see Fig. 3 discussion). The different symbols refer to the anneal temperature for each device. Mean values are shown for 140 °C annealed (optimized) samples.

Fig 5 shows the comparisons of PCB-esters with various R-groups, directly comparing efficiency ($\eta$), open circuit voltage ($V_{OC}$) and lifetime ($\tau$) for all the devices we have produced with these fullerene esters (nominal mole fraction $x=0.25$). Overall, the open-circuit voltages are comparable for the different esters, but there are some notable differences in both efficiency and lifetime.

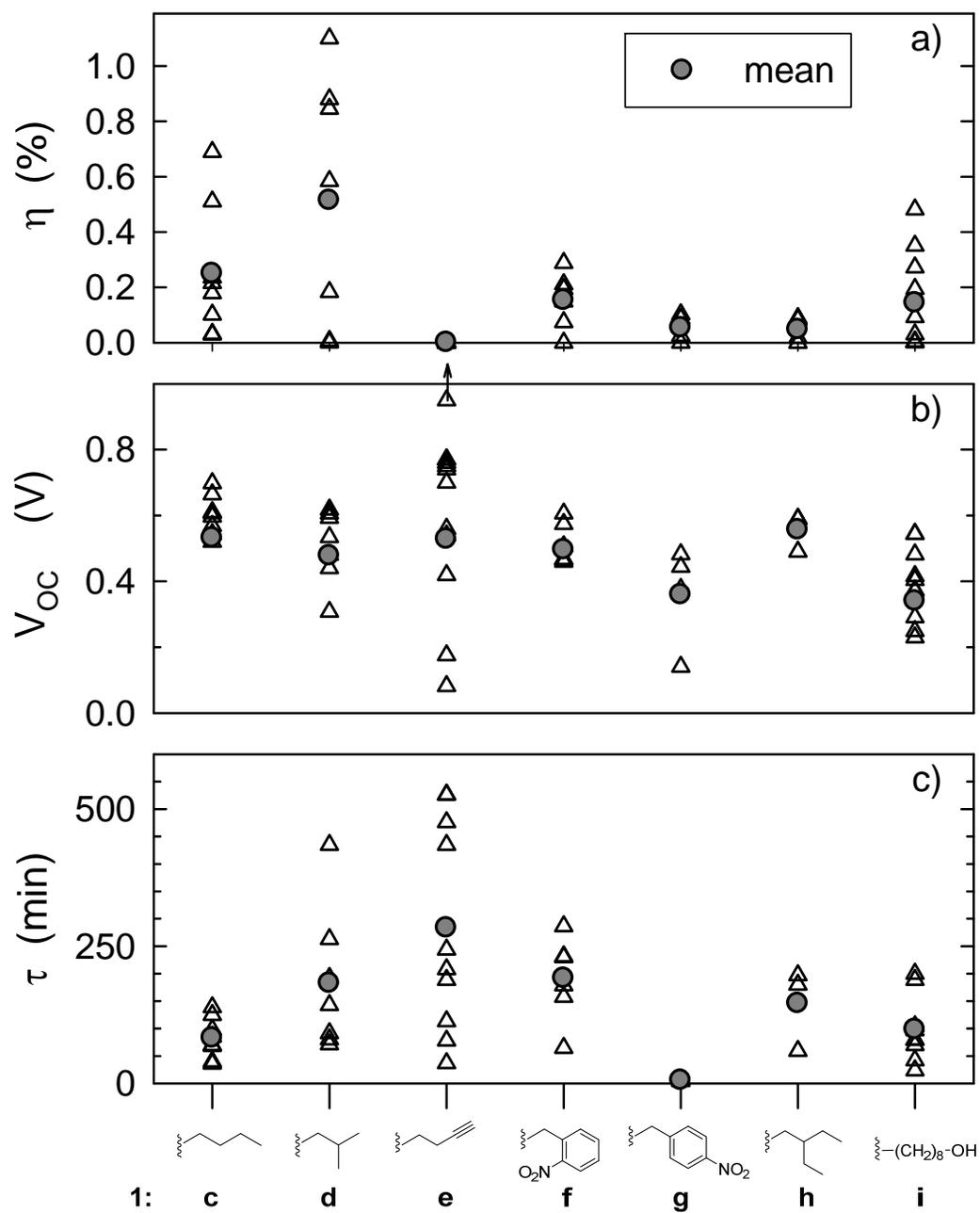

Fig. 5. Figures of merit for seven different fullerene structures, PCB-esters **1c-i**. Plotted are a) the initial power conversion efficiency $\eta$, b) the open circuit voltage $V_{OC}$ and c) the decay time constant $\tau$ for the fullerenes with esters indicated along the abscissa. The overall average values for each measurement are displayed as the gray circles.

The most surprising aspect of these results is that modest changes in R-group structure can have a dramatic affect on efficiency. The most significant change is the isomeric butyl vs. isobutyl esters, **1c** vs. **1d**, with the isobutyl being approximately three fold more efficient than butyl. Interestingly, isobutyl has a longer lifetime as well, which contradicts our previous findings with PCBM vs. PCBOD, in which we see a trade-off between efficiency and lifetime. The efficiency of butyl and isobutyl ester compared with the butynyl ester **1e** is also unexpected in that the ester chain length is the same and the alkyne functionality does not dramatically change the polarity of the side chain (or add a chromophore) and yet the performance of the device is dramatically affected with a large decrease in efficiency for the butynyl **1e**. The 2-ethylbutyl ester **1h** results correspond with the butynyl result in that a small change in structure (the addition of two carbons compared to butyl or isobutyl, but similar branching as the isobutyl) can dramatically decrease the efficiency of the device. The octanol-substituted ester **1g** shows a two-fold decrease in efficiency as compared to PCBO and about a ten-fold decrease in lifetime. The addition of the hydroxyl functionality does change the polarity/solubility of **1i** as compared to is closest structural analog PCBO, **1b** (**1i** has appreciable solubility in methanol, PCBO is not soluble in methanol). In contrast to the other results in this series, the changes we observe in efficiency and lifetime for **1i** are consistent with expectation that the more significant structural change would correspond to larger changes in device performance. This highlights an important feature of these results in that unexpected results can be obtained by examining "minor" structural modifications of the components in organic PV's. In our results, the magnitude of the change in efficiency and lifetime is roughly comparable for the fullerene esters we studied, yet the structural change is dramatically different; butyl vs. isobutyl or butyl vs. butynyl as compared to octyl vs. octyl-8-ol or octyl vs. octadecyl.

The isomeric nitrobenzyl esters, **1f** and **1g**, are especially interesting in that these isomeric compounds have dramatically different lifetimes. In fact, the 4-nitrobenzyl ester-based device decays faster than any of our previously studied fullerene ester devices (time constants of roughly 6 minutes). Again, such a dramatic change is unexpected given such a simple structural modification. One of our interests in the 2-nitrobenzyl ester, **1f**, was to use this ester as a photocleavable group that would allow us to study the fullerene acid in these devices by deprotecting the ester to form the carboxylic acid after device fabrication. Although ester cleavage could be an explanation for such a dramatic change in the device lifetime for the 4-nitrobenzyl ester, the 2-nitrobenzyl is the more photochemically labile yet has a comparable lifetime to the other fullerene esters we have studied. Although we have not ruled out photochemical reaction of **1g** as the cause of the rapid degradation, it is curious that the 2-nitrobenzyl ester does not exhibit the same behavior.

Fig. 6 shows a comparison between the efficiency of the 2-nitrobenzyl ester **1f** and 4-nitrobenzyl ester **1g**. Three separate samples of each are presented. A 6 minute lifetime curve (straight line) is shown for reference. We note that although the 4-nitrobenzyl ester **1g** exhibits a very rapid decay in efficiency, that effect is followed by a much slower ($\tau$ of roughly 300 minutes). This longer decay time is comparable to that of the 2-nitrobenzyl ester **1f**.

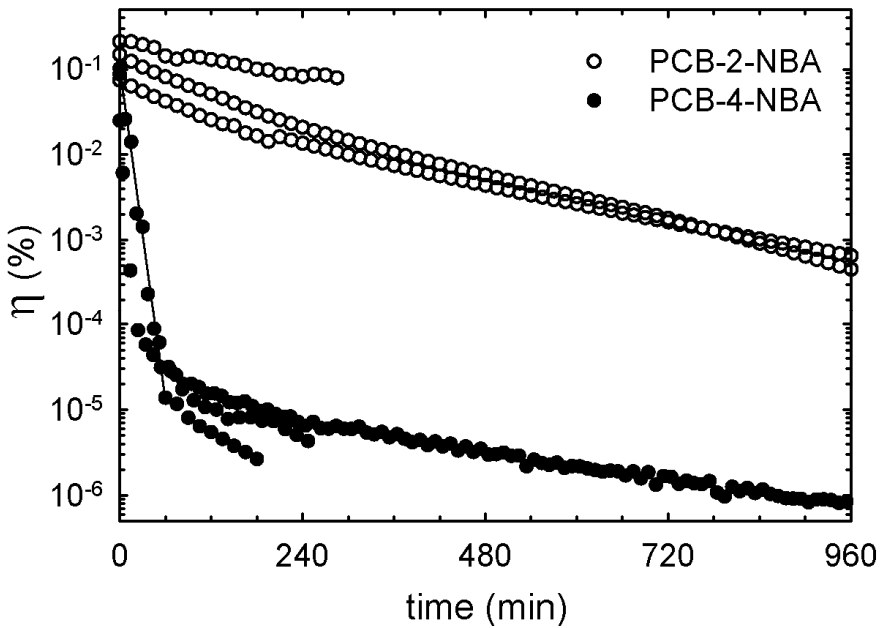

Fig. 6. Power conversion efficiency vs. time for 2-nitrobenzyl ester **1f** and 4-nitrobenzyl ester **1g**. Note the initially rapid degradation of the 4-nitrobenzyl ester followed by a much slower rate consistent with that of the 2-nitrobenzyl ester.

In our spectroscopic studies with PCBOD devices, we have established that the octadecyl chain has an effect on the polymer organization, decreasing the degree of crystallinity of the P3HT in the device and thus the efficiency of the device [23]. Since we observe that butyl and isobutyl ester chains are the highest efficiency devices in this series, with different efficiencies and lifetime figures of merit, we used our time-dependent UV-vis spectroscopic method to further characterize these film mixtures. The spectra of each and the changes developing over time are not significantly different from each other and are comparable to what we have observed with PCBM films [23]. The rate of change in the UV-vis spectra for each film maintains the same trend observed for transport results from device measurements for these two films.

In examining all of our spectral data, there are two regimes in which shorter chain fullerene ester-based films (octyl, butyl and isobutyl) produce UV spectra that are similar to the PCBM-based films but significant changes in ester chain length (octadecyl) produce films that are significantly thinner and absorb weakly in the 450-600 nm range, critical to device efficiency [23]. These results provide some broad guidelines with respect to fullerene ester structure selection. For device screening purposes, it would be useful to have a simple UV-vis "assay" that would be indicative film's efficiency but the spectroscopic changes over have time have closely correlated to the trends we observe in device lifetime measurements. As we gain more experience with the spectroscopy of these films, we are observing trends for higher performing devices although it is difficult to correlate smaller changes in efficiency to any specific spectroscopic changes for most of the esters we have studied.

The time-dependent spectroscopic results for the films containing both nitrobenzyl esters are consistent with the long time constant results observed in device studies (see Fig.6), however we were unable to observe any spectroscopic evidence for the rapid change in the 4-nitrobenzyl ester **1g**. Fig 7. Shows typical UV-vis data. A notable feature that was ubiquitous in our data for was a kink or pause in the spectral changes after about 1400-1500 minutes has elapsed (1 day). This behavior occurred in almost all samples, including a similar effect for a sample that was aged in the inert atmosphere glove box. This latter result suggests a morphological or structural transition is occurring, since chemical transformation due to oxidation should be suppressed in a nitrogen atmosphere. Overall, the spectral evidence has served as a valuable characterization of tool for these fullerene ester films and the rate of spectroscopic change is consistent with the temporal trends in the transport measurements.

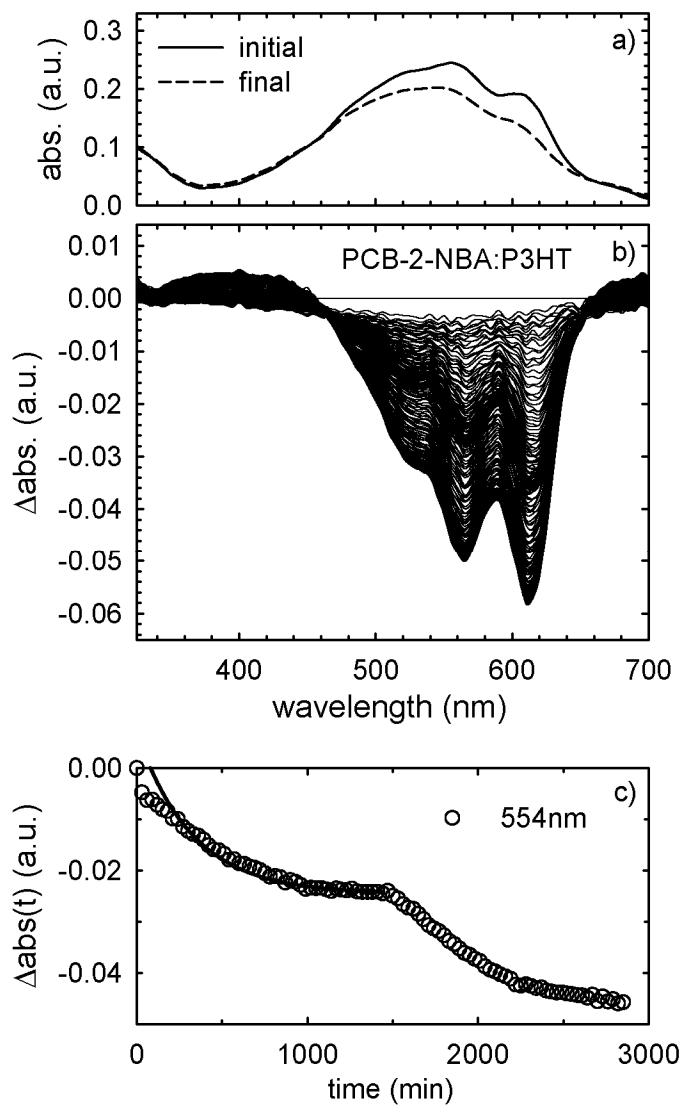

Fig.7. UV-vis absorption spectroscopy for a 2-nitrobenzyl ester, **1f**, sample: a) the initial and final absorbance, b) the differential absorbance taken at 15 minute intervals and c) the time dependence of the 554 nm absorption peak (every other data point removed for clarity). An exponential fit for the first 1500 minutes yields a time constant consistent with those derived from transport data (Fig. 6).

**Conclusions**

We have shown that unexpected findings are obtained by making minor perturbations in the structure of fullerene esters and observing the changes in the transport properties of solar cells. Our results show significant differences in the efficiency and lifetime of isomeric compounds in two different examples. Small structural changes such as and additional carbon or changes in the degrees of unsaturation also affect device performance. Although these results do not follow a "logical" pattern, the examination of small library of fullerene esters has clearly demonstrated the validity of using an array of compounds to determine structure-function relationships in organic photovoltaics. The library approach has been used to great effect in the development of pharmaceuticals, but has seen increasing use in materials research. We have demonstrated that this approach could be especially valuable in optimizing devices and that screening small libraries of compounds could yield the high performing devices required for field applications. Interestingly, this approach has recently been used to examine hole-conductor analogs in perovskite-based PV's where isomeric Spiro-OMeTAD analogs yielded devices with different efficiencies [30]. We conclude that applying this library approach would be a valuable tool in studying other thin-film photovoltaics and is an essential, yet infrequently used, method of device optimization.

We acknowledge both G. Laskowski and G. Sloan for invaluable technical assistance. Funding was provided by a Santa Clara University IBM Faculty Research Grant, a Santa Clara University Sustainability Grant and a grant from IntelliVision Technologies.

**Figure and Scheme Captions**

Fig. 1. Electron-donor and the two initial electron-acceptor molecules used in this study.

Scheme 1. Synthesis of fullerene esters **1a-i**

Fig. 2. Schematic of solar cell layers.

Fig. 3. Typical current-voltage (*I-V*) curves showing the degradation of a PCBO:P3HT device in ambient conditions. The arrow indicates the progression of time. *Inset:* a semilog plot of the power conversion efficiency of this device as a function of time in ambient conditions. The solid line fit shows the slope used to extract the characteristic time $\tau$.

Fig. 4. Efficiency and lifetime and data from a series of PCBO:P3HT devices. We plot a) the initial power conversion efficiency $\eta$ and b) the decay time constant $\tau$ as a function of PCBO mole fraction (see Fig. 3 discussion). The different symbols refer to the anneal temperature for each device. Mean values are shown for 140 °C annealed (optimized) samples.

Fig. 5. Figures of merit for seven different fullerene structures, PCB-esters **1c-i**. Plotted are a) the initial power conversion efficiency $\eta$, b) the open circuit voltage $V_{OC}$ and c) the decay time constant $\tau$ for the fullerenes with esters indicated along the abscissa. The overall average values for each measurement are displayed as the gray circles.

Fig. 6. Power conversion efficiency vs. time for 2-nitrobenzyl ester **1f** and 4-nitrobenzyl ester **1g**. Note the initially rapid degradation of the 4-nitrobenzyl ester followed by a much slower rate consistent with that of the 2-nitrobenzyl ester.

Fig.7. UV-vis absorption spectroscopy for a 2-nitrobenzyl ester, **1f**, sample: a) the initial and final absorbance, b) the differential absorbance taken at 15 minute intervals and c) the time dependence of the 554 nm absorption peak (every other data point removed for clarity). An exponential fit for the first 1500 minutes yields a time constant consistent with those derived from transport data (Fig. 6).